\documentclass[pdflatex,sn-mathphys-num]{sn-jnl}

\usepackage{graphicx}%
\usepackage{subcaption} 
\usepackage{multirow}%
\usepackage{amsmath,amssymb,amsfonts}%
\usepackage{amsthm}%
\usepackage{mathrsfs}%
\usepackage[title]{appendix}%
\usepackage{xcolor}%
\usepackage{textcomp}%
\usepackage{manyfoot}%
\usepackage{booktabs}%
\usepackage{algorithm}%
\usepackage{algorithmicx}%
\usepackage{algpseudocode}%
\usepackage{listings}%
\usepackage{amssymb}
\usepackage{xcolor}
\usepackage{soul}
\usepackage{url} 
\usepackage{natbib}

\usepackage{tabularx} 
\usepackage{booktabs} 

\theoremstyle{thmstyleone}%
\theoremstyle{thmstyletwo}%

\theoremstyle{thmstylethree}%

\begin{document}

\title[RAG]{Enhancing Health Information Retrieval with RAG by Prioritizing Topical Relevance and Factual Accuracy}


\author*[1]{\fnm{Rishabh} \sur{Upadhyay}}\email{uhrishabh@gmail.com}

\author[2]{\fnm{Marco} \sur{Viviani}}\email{marco.viviani@unimib.it}


\affil*[1]{\orgname{Inception AI}, \orgaddress{\city{Abu Dhabi}, \country{United Arab Emirates}}}

\affil[2]{
\orgname{University of Milano-Bicocca}, 
\orgaddress{\city{Milan}, \country{Italy}}}


\abstract{The exponential surge in online health information, coupled with its increasing use by non-experts, highlights the pressing need for advanced Health Information Retrieval models that consider not only topical relevance but also the factual accuracy of the retrieved information, given the potential risks associated with health misinformation. To this aim, this paper introduces a solution driven by Retrieval-Augmented Generation (RAG), which leverages the capabilities of generative Large Language Models (LLMs) to enhance the retrieval of health-related documents grounded in scientific evidence. In particular, we propose a three-stage model: in the first stage, the user's query is employed to 
retrieve topically relevant passages with associated references from a knowledge base constituted by scientific literature. In the second stage, these passages, alongside the initial query, are processed by LLMs to generate a contextually relevant rich text (GenText). 
In the last stage, the documents to be retrieved are evaluated and ranked both from the point of view of topical relevance and factual accuracy by means of their comparison with GenText, either through stance detection or semantic similarity. 
In addition to calculating factual accuracy, GenText can offer a layer of explainability for it, aiding users in understanding the reasoning behind the retrieval. Experimental evaluation of our model on benchmark datasets
and against baseline models demonstrates its effectiveness in enhancing the retrieval of both topically relevant and factually accurate health information, thus presenting a significant step forward in the health misinformation mitigation problem.}




\maketitle

\section{Introduction}\label{sec1}

\textcolor{black}{In the current online digital ecosystem, characterized by the problem of \textit{information overload} \citep{buchanan2001information}, health information is growing exponentially, and even non-expert users rely on it regularly. This underscores the critical need for sophisticated \textit{Information Retrieval} (IR) solutions for healthcare, able to address the specificities of this domain \cite{goeuriot2021clef}.} These systems, in fact, must not only ensure the retrieval of content that aligns with the intended topic, i.e., \textit{topical relevance}, but also rigorously uphold the principles of reliable and trustworthy information, i.e., \textit{factual accuracy} \cite{goodrich2019assessing}. \textcolor{black}{Actually}, in the realm of health, misinformation can have dire consequences, impacting individual decision-making, public health policies, and overall well-being \textcolor{black}{\cite{di2022health}}. 

Recently, there have been impressive advancements in \textcolor{black}{generative} \textit{Large Language Models} (LLMs) like \textcolor{black}{\textit{Generative Pre-trained Transformer}} (GPT)~\citep{abdullah2022chatgpt}, \textcolor{black}{\textit{Large Language Model Meta} AI} (Llama, formerly stylized as LLaMA) \citep{touvron2023llama}, and \textcolor{black}{\textit{General Language Model}} (GLM)~\citep{du2021glm} \textcolor{black}{as tools to carry out the most disparate tasks, including generating documents and answering specific questions in the healthcare sector \cite{ackerman2023automatic,frisoni2024generate,kell2024question}}. Although these models have shown remarkable general abilities in processing and generating human-like text~\citep{bang2023multitask,guo2023close}, their limitations in maintaining factual accuracy present significant hurdles~\citep{cao-etal-2020-factual,raunak-etal-2021-curious,10.1145/3571730}, in particular concerning the phenomena of \textcolor{black}{\textit{factual inconsistency} and} \textit{hallucination} \cite{saxena2023minimizing}. Factual inconsistency involves errors relative to known facts or logical coherence within the response. The model might provide wrong details (incorrect facts), distort facts or imply connections that do not exist in the source material (misleading summaries), or misunderstand the context and produce statements that are logically inconsistent with the given prompt (contextual errors). Hallucination involves creating content that has no basis in the real world. The model could generate plausible-sounding but entirely fictional information (fabricated information). 
This remains a critical issue that undermines the reliability of these tools \textcolor{black}{\cite{huang2023survey, zhang2023siren}}, particularly in fields requiring high precision such as legal advisement and healthcare. The rapid evolution of knowledge poses a challenge to the static datasets LLMs are typically trained on, resulting in outdated information~\citep{he2022rethinking}. The specificity and depth of knowledge required in specialized domains~\citep{li2023chatgpt,shen2023chatgpt} further exacerbate these models' limitations, as they often lack the nuanced understanding necessary for expert-level discourse and decision-making.

In response to these challenges, \textit{Retrieval-Augmented Generation} (RAG) has emerged as a compelling solution \cite{lewis2020retrieval} \textcolor{black}{to mitigate both factual inconsistency and hallucination \cite{perkovic2024hallucinations,saxena2023minimizing}}, 
by enhancing the precision and contextual relevance of responses generated by LLMs~\citep{borgeaud2022improving,guu2020retrieval,izacard2023atlas,lewis2020retrieval}. 
RAG, specifically, \textcolor{black}{increases the \textit{likelihood} of factual accuracy by integrating \textit{relevant} and \textit{trusted external knowledge} into the generation process of LLMs. First, it retrieves information that is relevant to the user's query from external knowledge sources (e.g., health ontologies, medical journal articles, etc.), which are more likely to provide factual and verified information. Next, LLMs are guided to generate their responses by focusing on the retrieved and reliable information, rather than solely depending on the model’s internal learned representations. This approach contributes to grounding the generation process in specific, relevant facts and domain-specific expertise \cite{setty2024improving,kang2024domain}.}


However, the integration of RAG into LLMs must be carefully managed to avoid potential pitfalls. \textcolor{black}{Practitioners should remain vigilant and not rely solely on the system's outputs. They should be empowered to monitor which information grounded in relevant facts effectively has been used to produce the final result. Indeed, although RAG is designed to reference authoritative knowledge bases, there remain risks if the system is not strictly limited to such reliable sources, or if those sources are ultimately not as reliable as expected. Indeed,} the vast and unfiltered nature of \textcolor{black}{Web-}based information introduces a risk of incorporating misinformation or `noise' if not properly curated, thereby potentially compromising the integrity of the responses 
\citep{adlakha2023evaluating}. These challenges necessitate a rigorous examination of the interaction between LLMs and RAG, particularly the extent to which Information Retrieval mechanisms are implemented to ensure they truly augment the model's performance by providing both (factually) accurate and (topically) relevant information.

\textcolor{black}{To take advantage of the benefits of RAG and in response to this last challenge, this paper proposes a RAG-driven IR model that capitalizes on the sophisticated text synthesis capabilities of generative LLMs and reliable external sources to enhance the retrieval of both topically relevant and factually accurate health-related documents. In particular,} our model strategically employs reputed medical journals as reliable sources of information,\footnote{\url{https://openmd.com/guide/finding-credible-medical-sources}} such as those accessible via the \textit{PubMed Central} (PMC) database,\footnote{\url{https://www.ncbi.nlm.nih.gov/pmc/}} a repository renowned for its comprehensive collection of validated health science literature, including the  \textit{Journal of the American Medical Association} (JAMA),\footnote{\url{https://jamanetwork.com/journals/jama}} and eLife.\footnote{\url{https://elifesciences.org/}} \textcolor{black}{At its core, the proposed solution utilizes a \textit{tripartite mechanism}.  First, we start with the extraction of passages from the above-mentioned external resources; such passages are topically relevant to user's queries and include bibliographic references. Subsequently, building upon these passages, our model leverages LLMs to generate a \textit{GenText}, i.e., contextually relevant rich tex constituted by explanatory and citation-rich responses, which serves as a basis for assessing the factual accuracy of health documents in the retrievable document collection. 
Finally, the ranking of documents is derived from the aggregation of topicality and factual accuracy scores, obtained by a basic IR model for topicality, and by performing stance detection and cosine similarity calculation between the \textit{GenText} and the document content for factual accuracy.} 


Through an extensive examination of the performance of the proposed solution on benchmark datasets from both the CLEF eHealth and TREC Health Misinformation tracks \cite{goeuriot2021clef,clarke2020overview}, we demonstrate its proficiency in \textcolor{black}{providing user access to relevant and likely accurate health information at the expense of likely health misinformation}. Also, by \textit{de facto} integrating explainability within \textit{GenText} \textcolor{black}{with scientific references}, we further \textcolor{black}{and briefly} illustrate how it could be possible to empower users to discern the underpinnings of the retrieved information \textcolor{black}{in the obtained ranking}, thus contributing to the fight against misinformation in health informatics. 

\textcolor{black}{Based on the aforementioned premises, and emphasizing that distinct technological solutions grounded in existing literature are adopted in a novel and combined manner within the proposed model, we summarize the main and original contributions of this work as follows:}
\begin{itemize}
    \item \textit{Integration of} RAG   \textit{for Health Information Retrieval}: The paper introduces a novel \textcolor{black}{IR} solution that combines RAG with advanced LLMs to 
    retrieve both topically relevant and factually accurate health information;
    \item \textit{Development of an enhanced factual accuracy \textcolor{black}{computation} method}: The paper proposes a novel solution to assess the factual accuracy of health information. This is achieved by generating \textit{GenText}, a contextually and evidence-rich textual representation to be compared against health information to be retrieved by means of stance detection and semantic similarity; 
    \item \textit{Improved explainability for users}: The insights derived from the enhanced factual accuracy assessment method 
    can be used to increase the explainability of the model. This makes the system more transparent for users retrieving health information through the proposed RAG-driven IR system, helping them understand the rationale behind the factual accuracy 
    of the obtained ranking.
\end{itemize}

\textcolor{black}{The remainder of the article is organized as follows: Section \ref{sec:relatedwork} discusses issues related to the identification of misinformation in general, how \textcolor{black}{recent solutions in IR have been} developed to take it—especially health misinformation—into account in the retrieval process, and the \textcolor{black}{current} solutions for misinformation detection based on RAG. Section \ref{sec:approach} focuses on describing the proposed model, which utilizes a RAG-driven model for Health Information Retrieval \textcolor{black}{based on \textit{GenText} and the proposed factual accuracy computation method}. Section \ref{sec:evaluations} is dedicated to experimental evaluations, discussing results and illustrating how \textit{GenText} could be used for the purpose of explainability of search results. Finally, Section \ref{sec:conclusions} outlines the conclusions and discusses future developments.}

\section{Related Work}\label{sec:relatedwork}

This work falls within the field of Information Retrieval, specifically \textit{Health Information Retrieval}, with the aim of \textcolor{black}{providing} users \textcolor{black}{with} relevant and factually accurate health information. 

In the literature, there are various strategies that attempt to combat (health) misinformation, often addressing the problem as a \textit{binary classification task} (i.e., information versus misinformation).
The most popular methods 
fall \textcolor{black}{mainly} into two categories: $(i)$ \textit{feature-based misinformation detection}, which involves training machine learning models on \textcolor{black}{distinct features extracted from the content (and related metadata) to be classified in terms of information/misinformation} \citep{zhao2023panacea, mendes2022human, yue2022contrastive, jiang2022fake, chen2023can, upadhyay2023vec4cred, 10.1145/3462203.3475898, 10.1145/3599696.3612902}, and $(ii)$ \textit{knowledge-based misinformation detection}, which involves gathering external knowledge to serve as corroborative evidence to validate the considered content \citep{brand2021bart, kou2022hc, wu2022bias, upadhyay2023explainable, shang2022knowledge}. This can include processing knowledge graphs or specific document fragments to support or dispute \textit{claims}. \textcolor{black}{In this context, the task of \textit{claim detection} is fundamental to ensuring the success of the content verification process against available knowledge bases \cite{hassan2017toward,guo2022survey}. Recently, there has also been an increasing focus on developing effective methods to place humans at the center of the misinformation identification process, \textcolor{black}{given the fact that automatic systems performing this task can still be subject to various forms of automation bias at different levels \cite{zeng2024combining}.}}


\textcolor{black}{Several of these approaches have been employed and tested within \textit{Information Retrieval Systems} (IRSs)\textcolor{black}{—a.k.a. search engines—}in recent years, also in relation to health to perform the task of \textit{Consumer Health Search} (CHS). CHS refers to the process of seeking health-related information by general consumers, typically through online search engines \cite{goeuriot2021consumer}. This encompasses a wide range of queries, from symptoms and treatments of illnesses to general health advice, diet, and wellness information. Unlike professional health search, which is conducted by healthcare providers using specialized databases and resources, CHS is performed by non-experts who may have varying levels of health literacy. The CLEF eHealth Lab Series,\footnote{\url{https://clefehealth.imag.fr/}} part of the \textit{Conference and Labs of the Evaluation Forum} (CLEF),\footnote{\url{http://clef-initiative.eu/}} has played a pivotal role in advancing research CHS. The goal of the initiative is to provide the research community with sophisticated datasets of clinical narratives, enriched with links to evidence-based care guidelines, systematic reviews, and other further information, to design ranking models considering multiple relevance dimensions such as \textit{topicality}, \textit{readability}, and \textit{credibility}, when retrieving documents w.r.t. user queries \citep{goeuriot2021consumer}.} \textcolor{black}{The TREC \textit{Heath Misinformation Track} \textcolor{black}{is an another initiative that specifically addresses the challenges of misinformation in health search.\footnote{\url{https://trec-health-misinfo.github.io/}} In particular, the goal of its \textit{ad-hoc retrieval} sub-task is to allow researchers working in the field} ``to design a ranking model that promotes \textit{credible} and \textit{correct} information over incorrect information'' \cite{clarke2020overview}.}

\textcolor{black}{
Among the recent research works submitted at CLEF,} in \textcolor{black}{\citep{di2020study} the authors evaluate \textit{Reciprocal Ranking Fusion} (RRF) \cite{cormack2009reciprocal} over different query variants, different retrieval functions, w/out
pseudo-relevance feedback, for both \textit{ad-hoc} and \textit{spoken queries retrieval} tasks, aiming to refine the relevance and readability of the retrieved information. The work proposed in \citep{mulhem2020lig} focuses on query expansion for \textit{ad-hoc retrieval} using the \textit{Unified Medical Language System} (UMLS)\footnote{\url{https://www.nlm.nih.gov/research/umls/index.html}} and the \textit{FastText} embedding model,\footnote{\url{https://fasttext.cc/}} putting a strong emphasis on enhancing terminological comprehensiveness. In \citep{seneviratne2020sandidoc}, the proposed solution utilizes TF-IDF scoring complemented by medical skip-gram word embeddings to experiment with different vector representations for textual data, aiming to optimize document-query similarity calculations. Concerning TREC submissions, 
models proposed by the CiTIUS Lab \citep{fernandez2020citius} and DigiLab \citep{zhang2022ds4dh} utilize BM25 as the basic IR model for ranking, complemented by sophisticated \textit{re-ranking} techniques employing RoBERTa and a combination of Transformer-based models to account for information credibility. Additionally, the works described in \citep{schlicht2021upv} and \citep{abualsaud2021uwaterloomds} leverage advanced \textit{Natural Language Processing} (NLP) tools including Bio-SBERT \citep{pankaj2022augmented} and T5 \citep{raffel2020exploring} models to refine search results, focusing on semantic similarity and the stance of documents.} Although these solutions have included formal semantic representations of texts that are often based on the use of Transformers, none of these have so far considered the use of generative LLMs to support the process of identifying and considering misinformation in the context of search engines.

\textcolor{black}{Instead, some work is increasingly appearing that uses generative LLMs to try to identify misinformation as a binary classification task \citep{wan2024dell,choi2024automated, cao2024can}, or to answer users' questions directly, including health-related questions \citep{wang2024jmlr, khlaut2024efficient}.}
\textcolor{black}{However, we are aware of the fact that} the knowledge \textcolor{black}{based on which such} LLMs \textcolor{black}{are trained is} commonly out-of-date~\citep{he2022rethinking}, and they also \textcolor{black}{risk to} generate factual inconsistent or hallucinated content, as previously introduced \citep{cao-etal-2020-factual,raunak-etal-2021-curious,10.1145/3571730}. To address these issues, current methodologies increasingly rely on \textit{Retrieval-Augmented Generation} (RAG) approaches ~\citep{borgeaud2022improving,guu2020retrieval,izacard2023atlas,lewis2020retrieval,shi2023replug,ren2023investigating}. These approaches enhance LLM responses by integrating retrieved external data, thus conditioning the generation process to be more factual and contextually relevant. RAG models have achieved remarkable results in various tasks such as open-domain QA~\citep{izacard-grave-2021-leveraging,trivedi-etal-2023-interleaving,li-etal-2023-large}, dialogue~\citep{cai-etal-2019-skeleton,cai-etal-2019-retrieval,peng2023check}, domain-specific question answering~\citep{cui2023chatlaw}, and code generation~\citep{zhou2023docprompting}. 
The LLM-Augmenter system presented in \citep{peng2023check} incorporates external knowledge and automated feedback mechanisms through plug-and-play modules to refine LLM output. \textcolor{black}{The authors in} \citep{chern2023factool} introduce a factuality detection framework that evaluates the authenticity of LLM-generated content across various tasks and domains. \textcolor{black}{The work described in} \citep{pan2023fact} leverages in-context learning capabilities of LLMs, employing \textit{Chain-of-Thought} (CoT) reasoning to guide models through complex problem-solving sequences. \textcolor{black}{In} \citep{zhang2023towards}, a \textit{Hierarchical Step-by-Step} (HiSS) prompting methodology \textcolor{black}{is proposed}, which systematically breaks down a claim into manageable sub-claims. This method sequentially verifies each sub-claim using question-answering techniques, relying on Web-retrieved evidence to ensure the factual integrity of responses. Concurrently, \citep{zeng2024justilm} emphasizes the necessity for generating sophisticated justifications for fact-checking claims, proposing a novel approach that focuses on the retrieval of evidence to support or refute claims rather than mere summarization.

Recently, \textcolor{black}{despite} a series of retrieval-enhanced tools and products have gained widespread attention, such as the ChatGPT retrieval plugin,\footnote{\url{https://github.com/openai/chatgpt-retrieval-plugin}} 
New Bing,\footnote{\url{https://www.bing.com/}} etc., \textcolor{black}{the technical details of these approaches are proprietary, nor is it publicly available how they effectively address the issue of hallucinations and other disadvantages related to the lack of factual accuracy. This is particularly concerning in the context of health misinformation, which is our primary focus of investigation. Therefore, we propose the solution illustrated in the following section.}

\section{\textcolor{black}{A RAG-Driven Model for Health Information Retrieval}}\label{sec:approach}

\begin{figure}[h!]
\caption{The three stages underlying the proposed RAG-driven model for Health Information Retrieval.}\label{fig:approach}
\centering
\includegraphics[width=\textwidth]{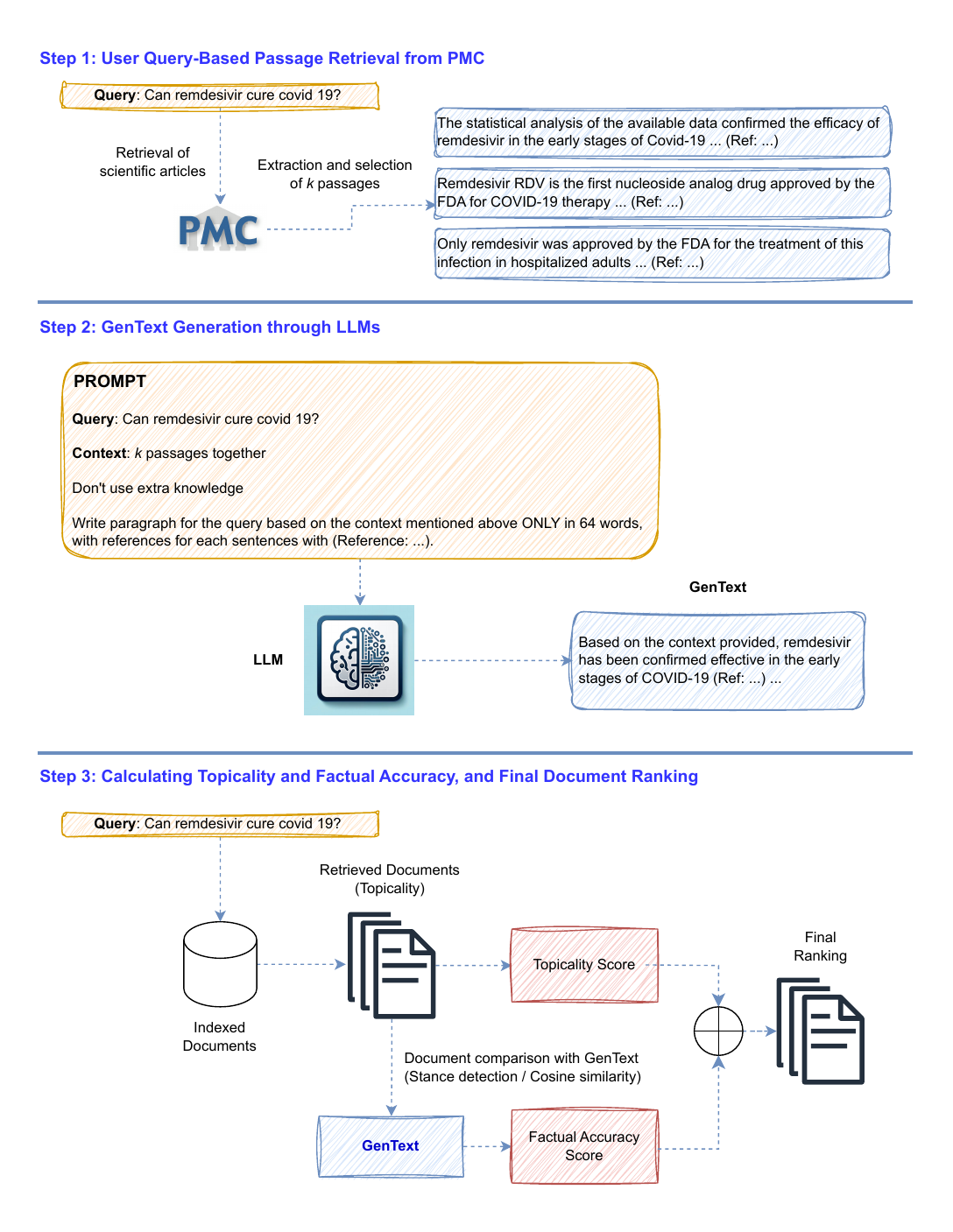}
\end{figure}

In this section, we delve into the proposed methodology underlying our RAG-driven model for Health Information Retrieval. As previously illustrated, central to our solution is the integration of generative LLMs with the curated scientific repository of PubMed Central (PMC), a strategy designed to \textcolor{black}{increase} both the topical relevance and factual accuracy of the retrieved documents. In particular, our solution is characterized by three key stages: $(i)$ \textit{user query-based passage retrieval from} PMC, $(ii)$ \textit{GenText generation through} LLMs, and $(iii)$ \textit{calculating topicality and factual accuracy, and final document ranking}. These three stages are graphically \textcolor{black}{illustrated in Figure \ref{fig:approach}\textcolor{black}{, and further detailed in the following sections}.}

\subsection{User Query-Based Passage Retrieval from PMC}


\textcolor{black}{The initial stage of our approach serves as retrieving top-$k$ scientific \textit{passages} from PMC that are \textit{topically relevant} with respect to the \textit{user's query}. It is constituted by further distinct steps, detailed in the following.}
\begin{itemize}
    \item In the first step, we leverage the PMC API to systematically retrieve \textit{scientific articles} from the PMC database in response to a given user's query;\footnote{\url{https://www.ncbi.nlm.nih.gov/home/develop/api/}} this is performed by employing a TF-IDF representation for both the query and the documents and the classical BM25 sparse retrieval model \cite{robertson2009probabilistic};
    \item Subsequently, the retrieved scientific articles are segmented into smaller \textit{passages}; in this work, each passage is constituted by one \textit{sentence};\footnote{\textcolor{black}{As noted by \citep{upadhyay2023passage}, models that operate at the sentence granularity typically perform better in terms of retrieving topically relevant and factually accurate consumer health information.}}
    \item At this point, the user's query and the retrieved passages are formally represented by \textit{contextualized word embeddings}; in particular, we employ \textcolor{black}{BioBERT} \citep{lee2020biobert}, specifically chosen for domain-adaptation;
    \item Following this, we calculate the \textit{cosine similarity} between the embeddings of the query and each passage. This similarity metric serves as an indicator of the topical relevance of each passage to the user's query. To further refine the retrieval process, we employ a \textit{discount scheme} where passages receive lower weights if they do not contain \textit{Named Entities} (such as specific medicines or diseases) that match those in the user's query 
    (Figure \ref{fig:query_retrieval_methods}).\footnote{This weighting strategy ensures prioritization of passages highly relevant to the user's query, particularly concerning specific medical terms and conditions, in the retrieval process.}

\begin{figure}[ht!]
    \centering
    \begin{subfigure}[b]{0.9\textwidth}  
        \includegraphics[width=0.8\textwidth]{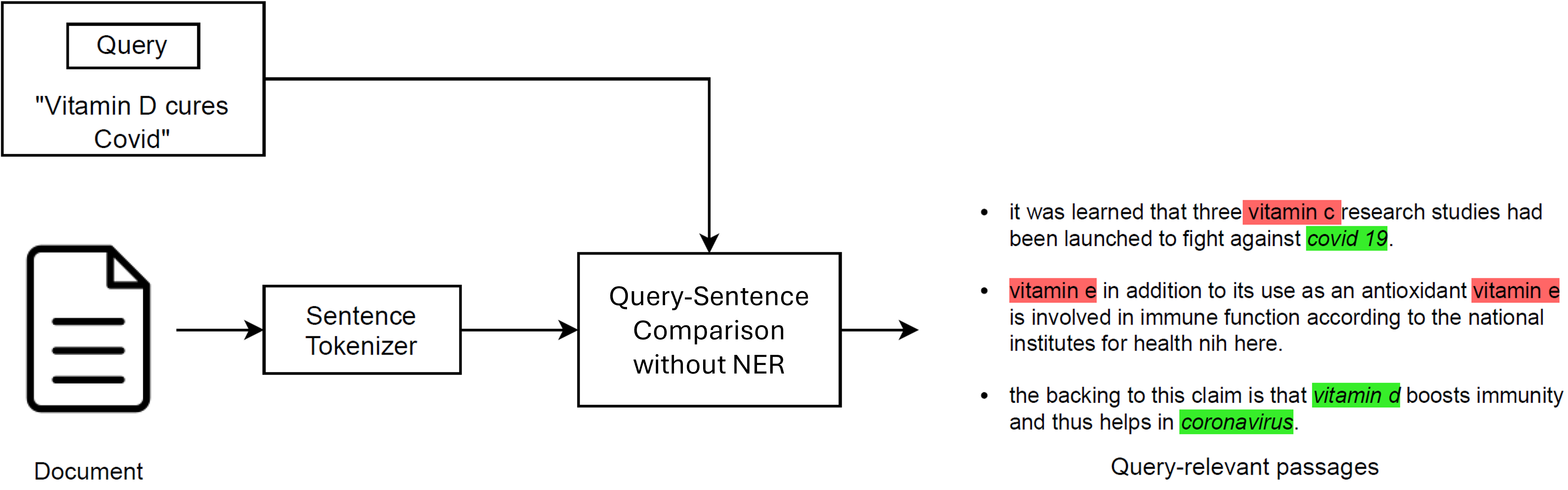}
        \caption{Query-based passage retrieval without NER.}
        \label{fig:passage_retrieval}
    \end{subfigure}
    \hfill  
    \begin{subfigure}[b]{0.9\textwidth}  
        \includegraphics[width=\textwidth]{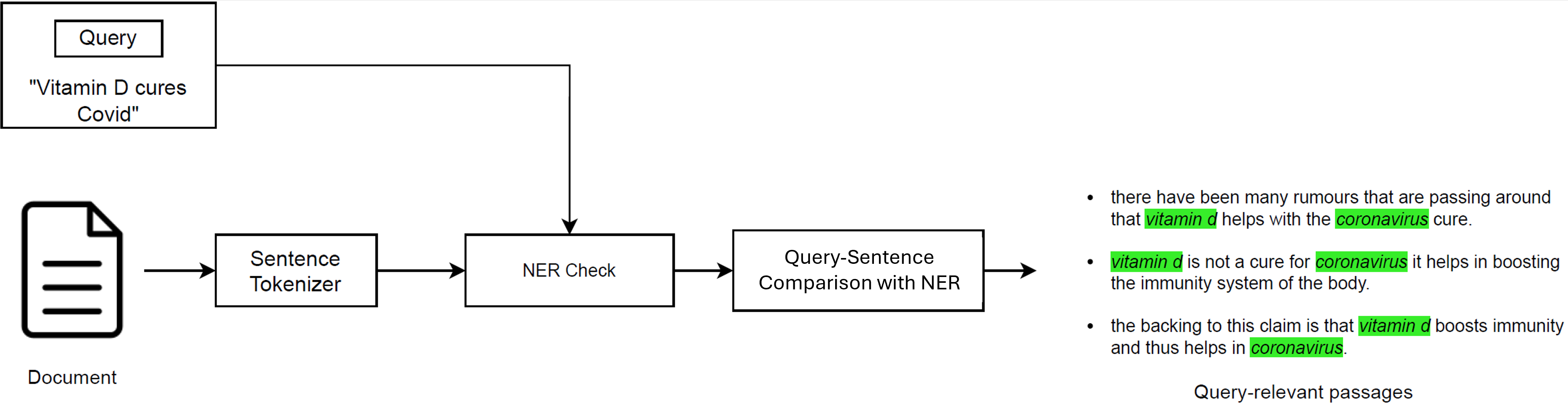}
        \caption{Query-based passage retrieval with NER.}
        \label{fig:NER_passage_retrieval}
    \end{subfigure}
    \caption{Comparison of query-based passage retrieval methods w/out NER.}
    \label{fig:query_retrieval_methods}
\end{figure}
    
    Specifically, the cosine similarity between each query $q$ and passage $p$ is initially computed to establish a preliminary ranking of passages based on topical relevance. This \textit{topicality score} is denoted as $sim(q,p)$. 
    At this point, if the query and the passage do not contain matching Named Entities referring to \emph{medicines} ($\mu$) and \emph{diseases} ($\delta$), the \textit{overall similarity score} $\sigma(q,p)$ is obtained by adjusting $sim(q,p)$ with a \textit{discount factor} $d_{\text{NE}}$ (where $d < 1$).\footnote{For finding the optimal $d_{\text{NE}}$ value, we used the method performed by \citep{upadhyay2023explainable}, i.e., a grid search using 5 queries (randomly selected) and document related to those queries. The grid search involved systematically testing different values of $d_{\text{NE}}$ within a predefined range, and evaluating the performance of the system for each value of $d_{\text{NE}}$ using a predefined metric (i.e., F1-score). The aim of this process was to identify the value of $d_{\text{NE}}$ that yielded the best performance in terms of the selected metrics, and therefore the best overall performance for the system.} Formally: 

    \begin{equation*}
        \sigma(q,p)=
    \begin{cases}
        sim(q,p),& \text{if } \mathrm{NER}_q(\mu,\delta)=\mathrm{NER}_p(\mu,\delta)\\
        d_{\text{NE}} \cdot sim(q,p),              & \text{otherwise}
    \end{cases}
    \end{equation*}
    \item It is important to emphasize that, as a final step, each retrieved passage is aligned with the \textit{bibliographic reference} of the original article from which it was extracted, denoted by its PMC ID. This alignment ensures the verifiability of the information sourced from the PMC database.
\end{itemize}

\subsection{GenText Generation through LLMs}

\textcolor{black}{By leveraging LLMs, given the user's query and the top-$k$ passages retrieved in the previous stage,\footnote{\textcolor{black}{Further detais about the selection of the optimal $k$ value are provided in Section \ref{sec:result}.}} we generate \textit{GenText}, i.e., a response text that answers the user's query including the relevant passages and citations. To do this, we define a \textit{prompt} guiding LLMs, which is constituted by:
\begin{enumerate}
    \item \textbf{Query:} The user's query;
    \item \textbf{Context:} The context of the query, i.e., the set of top-$k$ relevant scientific passages identified in the first stage;
    \item Additional prompt instructions.
\end{enumerate}
An example of a prompt instance used in our work is shown in Figure \ref{fig:llm_prompt}. In particular:}
\begin{figure}[!ht]
\centering
\footnotesize
\begin{tabularx}{\textwidth}{X}
\toprule
\textbf{LLM prompt} \\
\midrule
\textbf{Query}: can 5g antennas cause covid 19 \\
\\
\textbf{Context}: People around me told me not to get vaccinated against COVID-19 and reason 12 5G antennas are linked to the COVID-19 pandemic. At the same time there was no statistically significant difference in the average values of their answers regarding these reasons (Reference: 10316077).  Interference can have a significant impact on 5G networks particularly in the context of Internet of Things IoT devices. (Reference: 10144169)   These measures ensure that user privacy is protected and 5G networks can be trusted to handle massive data securely. The main causes and consequences of these challenges are summarized in Table 10 (Reference: 10255561).   The need to deal with the explosion of multimedia services has been considered in the 6G network which will provide greater QoS while also guaranteeing QoE (Reference: 10347022).  The importance of this was well proven in pandemic conditions of Covid-19 2729 So that in most organizations employees used different communication networks to do their work and after that the scope of communication networks in organizations has always grown. (Reference: 10399785)   Newly emerging variants of SARS-CoV-2 continue to pose a significant threat to global public health by causing COVID-19 epidemics (Reference: 10288941).  4 GHz transmit signal filters and amplifies the received signal and downconverts it to IF and then digitizes the signal according to the programmed parameter settings on the laptop (Reference: 9953371).   These tags incorporate antennas that can collect power efficiency regarding radio frequency queries from the RFID transceiver. Software is a good platform and crosses successful evolution for the sensor in Nanotechnology and bio-industries. In the medical field the Healthcare system is usually used to monitor the condition of patients (Reference: 10258751).  The controller of the level crossing must be equipped with antennas able to receive signals from the sensors in the train in a safe distance in the paper we suggested possible technologies of communication. The real-world application can be a mixed solution (Reference: 10384084).   The presence of various manufacturers and the diverse applications of sensors in disaster scenarios contribute to the heterogeneity of these sensors hence hindering the integration and sharing of information 107108. Some disasters may cause sequent disasters. For example, seismic activity or inundations can cause floods (Reference: 10490738). \\
\\
Write a paragraph answering the query based on the context provided above constituted by ONLY 64 words, with references for each sentence with (Reference:...). \\ \\
Do not use extra knowledge.\\
\bottomrule
\end{tabularx}\caption{Example of an instance of the prompt guiding LLMs.}\label{fig:llm_prompt}
\end{figure}
%
%
%
\begin{itemize}
    \item The additional prompt instruction: ``\textit{Write a paragraph answering the query based on the context provided above, constituted by ONLY 64 words, with references for each sentence (Reference:...)}'', has been meticulously designed to ensure the LLM's output is concise and information-rich. This word limit compels the LLM to synthesize the most relevant information efficiently; 
    \item The additional prompt instruction: ``\textit{Do not use extra knowledge}'', \textcolor{black}{aims at increasing the \textit{likelyhood} of} the LLM to confine its responses to the information provided in the context and not to incorporate any external details. \textcolor{black}{In fact, as discussed in both \citep{lewis2020retrieval} and \citep{izacard2020leveraging}, while these types of prompts can effectively reduce hallucinations and enhance factual alignment, they cannot completely eliminate the model's inclination to draw on prior knowledge from its pre-trained weights, given the current lack of a real introspection mechanism.}
\end{itemize}

\noindent The result of employing the prompt illustrated in Figure \ref{fig:llm_prompt} is as follows:
\begin{quote}
    \textit{Based on the context provided, there is a misconception linking 5G antennas to the COVID-19 pandemic (Reference: 10316077). However, this connection has no statistically significant evidence to support it (Reference: 10316077). Instead, it's important to note that 5G networks play a crucial role in ensuring secure data handling and enhancing user privacy (Reference: 10255561). Moreover, SARS-CoV-2 variants remain the main cause of COVID-19 outbreaks (Reference: 10288941).} 
\end{quote}

\subsection{\textcolor{black}{Calculating Topicality and Factual Accuracy, and Final Document Ranking}}\label{sec:topicalityfactualaccuracy}
\textcolor{black}{At this stage, documents in the document collection are retrieved based on both their \textit{topicality} and \textit{factual accuracy} relevance dimensions, in a \textit{multi-dimensional relevance} setting. 
Also this stage is constituted by distinct steps, detailed in the following.}
\begin{itemize}
    \item \textit{Calculating topicality}: The \textit{topicality score} $T$ is computed for each document $d$ in the document collection and it is derived using the BM25 retrieval model against the considered query $q$. Formally:  
    \[
    T(d,q) = \text{BM25}(d,q)
    \]
    \item \textit{Computing factual accuracy}: The \textcolor{black}{current proposal for computing the} \textit{factual accuracy score} $F$, integrates \textit{two distinct metrics} to assess the adherence of documents to scientific evidence within \textit{GenText} $G$. First, \textit{stance detection} is executed using the SCIFIVE \cite{phan2021scifive} model, a domain-specific T5 model pre-trained on large biomedical corpora \textcolor{black}{designed for biomedical text generation and understanding.}\footnote{\url{https://huggingface.co/razent/SciFive-large-Pubmed_PMC-MedNLI}} \textcolor{black}{In this paper, it is employed} to evaluate each document’s \textit{alignment} with the assertions in $G$\textcolor{black}{, considering such alignment an approximation toward factual accuracy.} This involves computing a so-called \textit{logits score}, denoted as \textit{stance}$(d, G)$, which provides a quantified measure of how much the \textit{stance} of document $d$ aligns with $G$. This score ranges between $0$ (indicating no support or contradiction) and $1$ (indicating maximum support). Additionally, we calculate a \textit{cosine similarity score}, denoted as \textit{cos}$(d, G)$, which measures the \textit{semantic similarity} between the embeddings of document $d$ and $G$. 
    The cosine similarity score also ranges between $0$ (no similarity) and $1$ (full similarity).
    \\
    The overall factual accuracy score $F$ for a document $d$ is then defined as a weighted combination of the stance detection and cosine similarity scores:
    \[
    F(d, G) = \alpha \cdot \text{\textit{stance}}(d, G) + (1 - \alpha) \cdot \text{\textit{cos}}(d, G)
    \]
    where $\alpha$ is a weighting factor used to balance the importance of stance detection \textcolor{black}{values} versus cosine similarity \textcolor{black}{values} in \textcolor{black}{computing} factual accuracy \textcolor{black}{as defined}. 
    \item \textit{Ranking documents}: The final document ranking is \textcolor{black}{obtained by performing a} \textit{linear combination} of topicality and factual \textcolor{black}{accuracy} scores \textcolor{black}{in order to obtain the \textit{Retrieval Status Value} (RSV) based on which the ranking is obtained}. 
    \textcolor{black}{Formally:}
     \[
    \text{RSV}(d, q, G) = \beta \cdot T(d, q) + (1 - \beta) \cdot F(d, G)
    \]
    In the above equation, $\beta$ is a weighting factor that allow us to balance the importance of topicality with respect to factual accuracy in the final document ranking.
\end{itemize}

\section{\textcolor{black}{Experimental Evaluation}}\label{sec:evaluations}

This section presents the results of the experimental evaluation conducted to assess the effectiveness of the RAG-driven IR model for Health Information Retrieval, in the context of Consumer Health Search. The data on which the evaluation was conducted are those made publicly available by the two previously introduced 2020 CLEF eHealth and TREC Health Misinformation evaluation initiatives.

\subsection{Datasets}
The CLEF eHealth 2020 dataset consists of \textit{Web pages} gathered by \textit{Common Crawl},\footnote{\url{https://commoncrawl.org/}} in the time interval 2018-2020, around 50 distinct medical conditions. The dataset has a fixed structure, organized into distinct \textit{scenarios} (i.e., distinct medical conditions). To each scenario are associated a \textit{query} and a \textit{narrative}, which contextualizes the query. For example, for the query: ``list of multiple sclerosis symptoms'', the associated narrative is: ``I am a 40 year old patient with MS, and I have very vague
symptoms, including fatigue, brain fog, foot drop, difficulties passing
urine, problems turning right. Are these related to MS or might I have another disease in addition?''. The data is labeled with respect to \textit{topicality}, \textit{understandability},
and \textit{credibility}. In particular, understandability is an estimation of whether the document is understandable by a patient, and credibility is the concept that, in the document collection, is used to approximate that of information truthfulness. In this work, we just considered as labels topicality and
credibility. Both of them are provided on a binary scale, i.e., topically relevant or non-topically relevant, and
credible or non-credible.

The TREC Health Misinformation 2020 dataset is constituted by \emph{CommonCrawl news}, 
sampled from January, 1st 2020 to April 30th, 2020, which contains health-related news articles 
encompassing 46 topics linked to Coronavirus. 
The dataset has a fixed structure, organized into \emph{topics}. Each topic includes a \emph{title}, a \emph{description}, which reformulates the title as a question, a \emph{yes/no answer}, which is the actual answer to the description field based on the provided evidence, and a \emph{narrative}, which describes helpful and harmful documents in relation to the given topic. For example, for the topic title field: ``ibuprofen COVID-19'', the value of the other attributes in the dataset are, for the description: ``Can ibuprofen worsen COVID-19?'', for the yes/no answer: ``no'', and for the narrative: ``Ibuprofen is an anti-inflammatory drug used to reduce fever and treat pain or inflammation''.
The data is labeled with respect to \textit{usefulness}, \textit{answer}, and \textit{credibility}. In particular, usefulness corresponds to topical relevance, answer indicates if the document provides an answer to the query contained in the description field, and credibility approximates information truthfulness. In this work, we just considered as labels usefulness and credibility. Both of them are provided on a binary scale, i.e., useful or non-useful, and credible or non-credible.

\subsection{Implementation Details}

In our study, we employed \textit{Python} 3.8 as the primary programming language due to its extensive libraries and ease of integration with various tools.\footnote{\url{https://www.python.org/downloads/release/python-380/}}

\textcolor{black}{As LLMs for the generation of \textit{GenText}, we considered GPT-3, Llama 3, and Falcon-40B. For GPT-3 inferencing, we used OpenAI’s API.\footnote{\url{https://openai.com/index/openai-api/}} For Llama 3 and Falcon-40B inferencing,} 
we used the \textit{Ollama} framework.\footnote{\url{https://ollama.com/}}

For NLP tasks \textcolor{black}{maily related to data preprocessing and word embedding}, we utilized the \textit{Hugging Face Transformers library}.\footnote{\url{https://huggingface.co/docs/transformers/index}} The \textit{data preprocessing} steps were crucial to ensure the quality and effectiveness of the training process. \textit{Text normalization} was performed as an initial step, wherein all textual data were converted to lowercase, punctuation was removed, and whitespace was stripped. This process helped in reducing variability and ensuring consistency across the dataset.
Subsequently, \textit{tokenization} was applied using the BERT tokenizer. This choice was made to maintain alignment with the input requirements of the BERT model, ensuring that the texts were appropriately segmented into tokens that the model could process efficiently. Tokenization is an essential step in NLP tasks as it transforms raw text into a structured format suitable for model consumption.

To facilitate the \textit{training process} and \textit{accelerate computations}, we utilized an NVIDIA Tesla A10 GPU. The choice of this hardware allowed for the efficient handling of large datasets and computationally intensive tasks, significantly speeding up the training phase.

\subsection{Baselines}

The performance of the proposed RAG-driven IR model was evaluated with respect to the fact that the model itself can be driven by different LLMs in generating \textit{GenText}. As anticipated earlier, we considered GPT, Llama, and Falcon, so the three model configurations will be referred to in the following as GPT$_{\text{RAG}}$, \textit{Llama}$_{\text{RAG}}$, and \textit{Falcon}$_{\text{RAG}}$. The performance of \textcolor{black}{the proposed model guided by} such LLM configurations has been compared to several \textcolor{black}{IR} \textit{baseline models} from the literature. Each of these models utilizes distinct methodologies designed to enhance IR \textcolor{black}{effectiveness} by incorporating various layers of semantic understanding, topical relevance, and factual accuracy assessment\textcolor{black}{, but none of them rely on the use of LLMs, nor RAG \textcolor{black}{(as previously illustrated in Section \ref{sec:relatedwork})}.}
\begin{itemize}
    \item BM25: \textcolor{black}{It is the state-of-the-art IR model} that acts as the \textit{foundational baseline} in our comparisons. This model accounts for \textit{topical relevance only} and evaluates documents by scoring them according to the frequency of query terms, with adjustments made for document length;
    \item DigiLab: The model proposed in \cite{zhang2022ds4dh} employs a \textit{re-ranking} approach that extends beyond the capabilities of BM25. Initially, BM25 is used to generate a preliminary set of potentially topically relevant documents. This is followed by a re-ranking stage utilizing a multi-model pipeline that evaluates documents on three key aspects: \textit{usefulness}, assessed by Transformer-based models fine-tuned on the MS MARCO dataset \cite{bajaj2016ms}; \textit{supportiveness}, evaluated by BERT models fine-tuned on scientific literature and Wikipedia; and \textit{credibility}, determined by a Random Forest model trained on the Microsoft Credibility dataset \citep{schwarz2011augmenting}. The results from these evaluations are then integrated using Reciprocal Rank Fusion, which enhances the relevance and credibility of the retrieved documents;
    \item CiTIUS: Also the model presented in \cite{fernandez2020citius} employs \textit{re-ranking}. Initially, BM25 ranks the top-100 documents according to their topical relevance to the query. The subsequent re-ranking involves a more in-depth analysis: RoBERTa is used for \textit{semantic representation}, encoding sentences to assess their similarity to the query topic. Additionally, a \textit{reliability classifier} trained on historical data evaluates the trustworthiness of these passages. Finally, scores from these components are combined using techniques like CombSUM or \textit{Borda Count}, producing the final document ranking;
    \item WISE: The model presented in \citep{upadhyay2022unsupervised} begins by retrieving documents using BM25 based on user queries. It then calculates a \textit{truthfulness score} for these documents by cross-referencing their claims with supporting scientific articles, using cosine similarity to measure alignment. Finally, WISE re-ranks the documents by linearly combining the BM25 scores with the truthfulness scores;
    \item WISE$_{\text{NLI}}$: Building on the WISE model, WISE$_{\text{NLI}}$ incorporates \textit{Natural Language Inference} (NLI) techniques to improve the evaluation of document truthfulness. In addition to cross-referencing scientific articles, it assesses the stance of document content relative to claims found in the referenced scientific literature. \textit{Stance scores} generated by NLI models quantify the degree of agreement or disagreement with established facts. The final document ranking is then refined by linearly combining BM25 scores, truthfulness scores, and stance scores.
\end{itemize}

\subsection{Experimental Results}
\label{sec:result}

The results of the model evaluations tested are shown in this section with respect to the two datasets considered in terms of CAM$_{\text{MAP}}$ and CAM$_{\text{NDCG}}$, two advanced IR evaluation metrics illustrated in detail in \citep{lioma2017evaluation}. It is important to note that these results refer to: the optimal number of retrieved scientific articles, \textcolor{black}{the optimal number $k$ of topically relevant passages to be retrieved from PCM, and the optimal} $\alpha$ and $\beta$ values related to the calculation of factual accuracy and RSV respectively, as shown in Section \ref{sec:topicalityfactualaccuracy}, as determined to maximize the system's performance.\footnote{\textcolor{black}{It was shown in \cite{upadhyay2023explainable} that excellent results are obtained by considering only the first highly relevant scientific article. The optimized values of the other three parameters were obtained by considering a subset of the data used for evaluating the proposed solution (5 queries from each dataset and their associated documents). The model's effectiveness on this subset was assessed using the previously described metrics, evaluating both topical relevance and factual accuracy to achieve the best balance between the two. 
Through this process, we were able to obtain $k=10$ (we tested from $k=5$ to $k=20$), $\alpha=0.65$, and $\beta=0.45$. Additionally, the queries used for parameter tuning were removed from the final testing to prevent overfitting and ensure a robust evaluation of the system on unseen data.}
}

\begin{table}[!h]
\centering
\caption{Performance comparison of baselines and model configurations on the CLEF eHealth 2020 dataset for the top-5 and top-10 retrieved documents.}
\label{tab:clef_performance}
\small
\begin{tabularx}{\textwidth}{Xccc}  
\toprule
\textbf{Model} & \textbf{CAM$_{\text{MAP}}$} & \textbf{CAM$_{\text{NDCG}}$} & \textbf{Embeddings} \\
\midrule
\multicolumn{4}{c}{\textbf{Top-5 Documents}} \\
\midrule
BM25 & 0.0431 & 0.1045 & - \\
DigiLab & 0.0433 & 0.1109 & - \\
CiTIUS & 0.0455 & 0.1119 & - \\
WISE & 0.0611 & 0.1198 & BioBERT \\
WISE$_{\text{NLI}}$ & 0.0883 & 0.1823 & BioBERT \\
GPT$_{\text{RAG}}$ & 0.1045 & 0.2098 & BioBERT \\
\textit{Llama}$_{\text{RAG}}$ & \textbf{0.1079} & \textbf{0.2146} & BioBERT \\
\textit{Falcon}$_{\text{RAG}}$ & 0.0994 & 0.2011 & BioBERT \\
\midrule
\multicolumn{4}{c}{\textbf{Top-10 Documents}} \\
\midrule
BM25 & 0.0784 & 0.1923 & - \\
DigiLab & 0.0823 & 0.1992 & - \\
CiTIUS & 0.0843 & 0.1999 & - \\
WISE & 0.1102 & 0.211 & BioBERT \\
WISE$_{\text{NLI}}$ & 0.1302 & 0.2321 & BioBERT \\
GPT$_{\text{RAG}}$ & 0.1502 & 0.2655 & BioBERT \\
\textit{Llama}$_{\text{RAG}}$ & \textbf{0.1532} & \textbf{0.2702} & BioBERT \\
\textit{Falcon}$_{\text{RAG}}$ & 0.1495 & 0.2568 & BioBERT \\
\bottomrule
\end{tabularx}
\end{table}


\begin{table}[!h]
\centering
\caption{Performance comparison of baselines and model configurations on the TREC Health Misinformation 2020 dataset for the top-5 and top-10 retrieved documents.}
\label{tab:trec_performance}
\small
\begin{tabularx}{\textwidth}{Xccc} 
\toprule
\textbf{Model} & \textbf{CAM$_{\text{MAP}}$} & \textbf{CAM$_{\text{NDCG}}$} & \textbf{Embeddings} \\
\midrule
\multicolumn{4}{c}{\textbf{Top-5 Documents}} \\
\midrule
BM25 & 0.0631 & 0.1435 & - \\
DigiLab & 0.0712 & 0.1543 & - \\
CiTIUS & 0.0754 & 0.1554 & - \\
WISE & 0.0844 & 0.1608 & BioBERT \\
WISE$_{\text{NLI}}$ & 0.0923 & 0.1922 & BioBERT \\
GPT$_{\text{RAG}}$ & 0.1178 & 0.2234 & BioBERT \\
\textit{Llama}$_{\text{RAG}}$ & \textbf{0.1222} & \textbf{0.2298} & BioBERT \\
\textit{Falcon}$_{\text{RAG}}$ & 0.1123 & 0.2165 & BioBERT \\
\midrule
\multicolumn{4}{c}{\textbf{Top-10 Documents}} \\
\midrule
BM25 & 0.1047 & 0.2052 & - \\
DigiLab & 0.1186 & 0.2011 & - \\
CiTIUS & 0.1194 & 0.2095 & - \\
WISE & 0.1233 & 0.22 & BioBERT \\
WISE$_{\text{NLI}}$ & 0.1341 & 0.2455 & BioBERT \\
GPT$_{\text{RAG}}$ & 0.1547 & 0.2712 & BioBERT \\
\textit{Llama}$_{\text{RAG}}$ & \textbf{0.1602} & \textbf{0.2723} & BioBERT \\
\textit{Falcon}$_{\text{RAG}}$ & 0.1501 & 0.2665 & BioBERT \\
\bottomrule
\end{tabularx}
\end{table}


The results illustrated in both Table \ref{tab:clef_performance} and Table \ref{tab:trec_performance} highlights the substantial improvements achieved by RAG-driven model configurations with respect to the considered baseline models. Traditional models like BM25 and more advanced systems such as DigiLab and CiTIUS provide foundational benchmarks, showing basic to moderate improvements in retrieval metrics. However, models that integrate knowledge bases and evidential reasoning, such as WISE and WISE$_{\text{NLI}}$, demonstrate significant performance enhancements, emphasizing the crucial role of semantic processing in improving document topical relevance and factual accuracy. However, the standout performers are the RAG-driven model configurations, which consistently excel in both the top-5 and top-10 document retrieval categories, especially with regard to the RAG model led by the Llama LLM. These models effectively leverage deep language understanding to deliver both contextually relevant and factually accurate results. 

\subsection{Enhancing Explainability with GenText}\label{sec:explainability}

In this section we briefly illustrate the possibility of using \textit{GenText} as a means of increasing the explainability of the obtained search results, and provide an example.
%
%
Indeed,
the text generated by the LLM is crafted to answer user queries by providing concise, factual information backed by citations from scientific sources. Such information can be provided along with the search results to make it clearer to users which sources are reliable against which such documents have been judged factually accurate.

\begin{figure}[h!]
\centering
\includegraphics[width=1\linewidth]{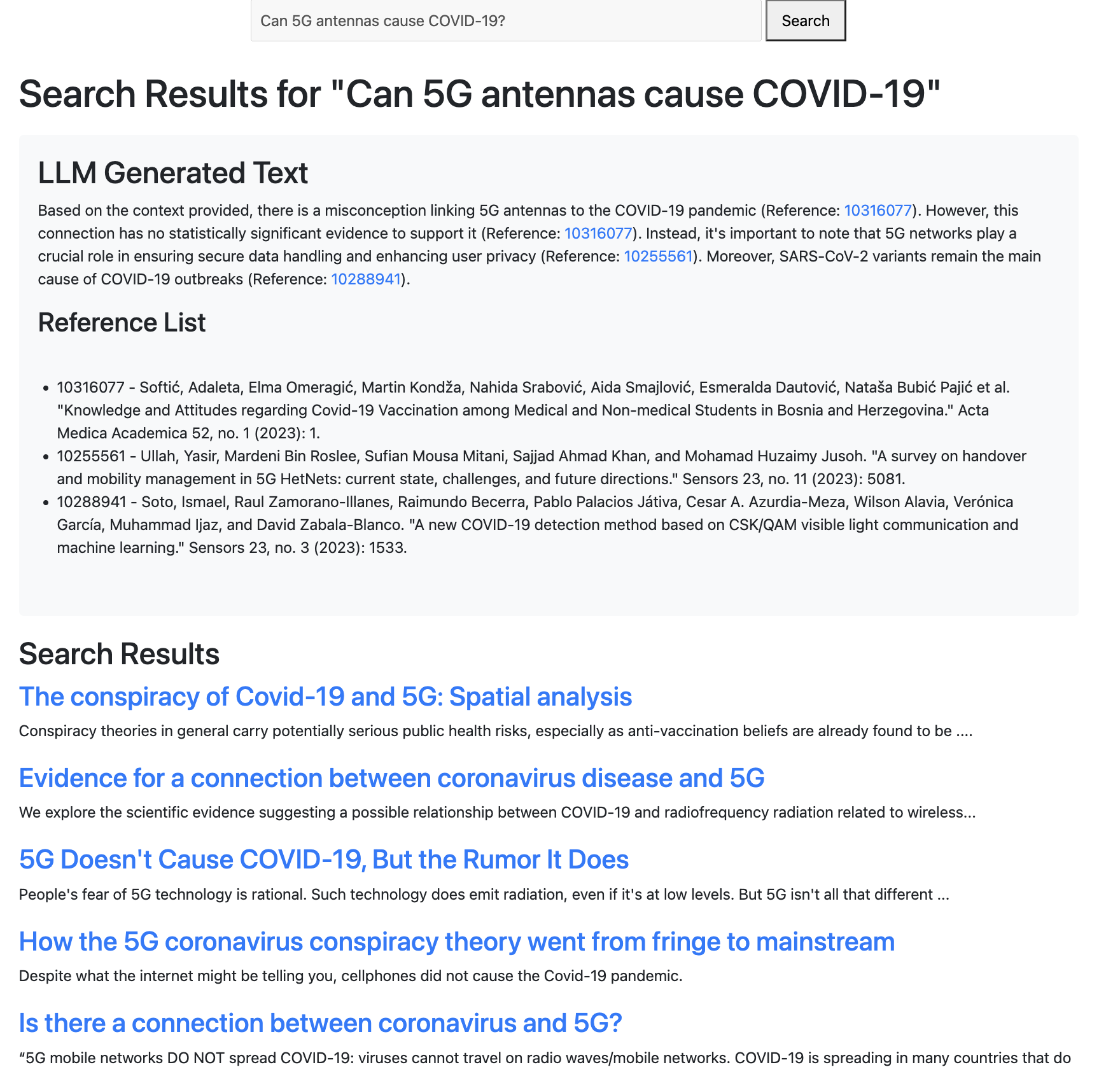}
\caption{Example of \textit{GenText} used for explainability, addressing a query about the relationship between 5G technology and COVID-19 with references to scientific literature.}
\label{fig:ex}
\end{figure}

For example, in Figure \ref{fig:ex}, the query ``Can 5G antennas cause COVID-19?" is answered with a explanation that highlights the absence of any scientific evidence linking 5G technology with the COVID-19 pandemic, thereby addressing a misinformation. The response not only refutes the misinformation but also enriches the user's understanding by referencing relevant scientific literature that supports the facts.


\section{Conclusion and Discussion}\label{sec:conclusions}

In this paper, we have introduced a novel approach to Health Information Retrieval through a Retrieval-Augmented Generation (RAG) model, which aims to tackle the challenge of effectively accounting for both topical relevance and factual accuracy in the retrieval process. The exponential growth of online health information and its widespread consumption by non-experts highlight, in fact, the urgent need for robust models capable of mitigating the risks associated with incurring online health misinformation. To this aim, our proposed three-stage model harnesses the strengths of generative Large Language Models (LLMs) to improve the retrieval of health-related documents that are both scientifically grounded and contextually relevant.
The experimental results obtained illustrate that our RAG-driven model outperforms existing baseline models in producing ranked results that account for both dimensions of relevance. This enhancement relies in particular on our model's ability to generate GenText, a rich and contextually aware text that serves not only as a benchmark for evaluating factual accuracy—an additional innovative aspect of our approach—but also as a valuable tool for promoting explainability.

\textcolor{black}{The aspect of explainability is of particular interest in a model like the one proposed in this work. In fact, while our model notably enhances the balance between topicality and the likelihood of factual accuracy in the final ranked list of documents, it is essential to recognize that a potential risk of automation bias remains, linked to various factors. On one hand, RAG does not guarantee that LLMs will produce content that aligns solely with trusted knowledge bases. Moreover, there are limitations to our solution; for instance, the assessment of factual accuracy provides an approximation rather than a definitive measure of truth. This leaves ample room for future research, both in exploring introspection and reasoning mechanisms within LLMs and in investigating alternative solutions for calculating a value of factual accuracy to be incorporated into the RAG-based IR model.}
\textcolor{black}{In addition, while our study has focused on generating GenText using three general-purpose LLMs—GPT, LLaMA, and Falcon—it would be beneficial to evaluate the application of domain-specific LLMs, such as Med-PaLM, optimized for answering clinical questions; MedMT5, which excels at multilingual generation tasks; and DocOA, which integrates RAG techniques specifically for personalized osteoarthritis management, demonstrating the potential for tailored medical recommendations. Moreover, considering general-purpose models fine-tuned on medical literature could offer additional insights into optimizing factual accuracy and relevance in Health Information Retrieval.}

\backmatter






\section*{Declarations}

\noindent \textbf{Author contribution.} All authors contributed to the study's conception and design. Material preparation, data collection, and analysis were performed by Rishabh Upadhyay. The first draft of the manuscript was written by Rishabh Upadhyay. Marco Viviani commented on previous versions of the manuscript. All authors read and approved the final manuscript.\\

\noindent \textbf{Conflict of interest.} The authors declare they have no financial interests.\\

\noindent\textbf{Data availability.} All data analysed during this study are included in the following published articles and their supplementary information files: \cite{goeuriot2021clef,clarke2020overview}.\\

\noindent \textbf{Ethical approval.} Not applicable.\\

\noindent \textbf{Funding.} This work was supported by the Italian Ministry of University and Research (MUR) under PRIN 2022 project KURAMi: ``Knowledge-based, explainable User empowerment in Releasing private data and Assessing Misinformation in online environments'' (20225WTRFN).\footnote{\url{https://kurami.disco.unimib.it/}}

\bibliography{sn-article}

\end{document}